\PassOptionsToPackage{svgnames}{xcolor}
\documentclass[galaxies,review,accept,pdftex,moreauthors]{Definitions/mdpi} 

\firstpage{1} 
\pubvolume{13}
\issuenum{4}
\articlenumber{81}
\pubyear{2025}
\copyrightyear{2025}
\datereceived{25 March 2025} 
\daterevised{3 July 2025} 
\dateaccepted{14 July 2025} 
\datepublished{17 July 2025} 
\hreflink{https://doi.org/galaxies13040081} 

\usepackage{multirow}

\def\msol{\ensuremath{M_\odot}}
\def\lsol{\ensuremath{L_\odot}}
\def\rsol{\ensuremath{R_\odot}}

\def\teff{\ensuremath{T_\text{eff}}}

\def\hsg{Hertzsprung gap}
\def\donc{\ensuremath{\ \rightarrow\ }}

\Title{Stellar evolution through the Red Supergiant phase}

\TitleCitation{Stellar evolution through the Red Supergiant phase}

\Author{Sylvia Ekstr\"om $^{1,*,\dagger}$\orcidA{} and Cyril Georgy $^{1,\dagger}$\orcidB{}}

\AuthorNames{Sylvia Ekstr\"om and Cyril Georgy}

\AuthorCitation{Ekstr\"om, S.; Georgy, C.}

\address{%
$^{1}$ \quad University of Geneva, Department of Astronomy, Chemin Pegasi 51, 1290 Versoix GE, Switzerland}

\corres{Correspondence: sylvia.ekstrom@unige.ch}

\firstnote{These authors contributed equally to this work.}

\abstract{Massive stars less massive than $\sim$30\,{\msol} evolve into a red supergiant after the main sequence. Given a standard IMF, this means about 80\% of all single massive stars will experience this phase. RSGs are dominated by convection, with a radius that may extend up to thousands of solar radii. Their low temperature and gravity make them prone to lose large amounts of masses, either through a pulsationally-driven wind or through mass-loss outburst. RSGs are the progenitors of the most common core-collapse supernovae, the type II. In the present review, we give an overview of our theoretical understanding about this spectacular phase of massive stars evolution.}


\begin{document}
\section{Introduction}
The red supergiant (RSG) stage occurs after the main sequence (MS) of massive stars between $\sim$9-30\,{\msol}. Their direct progenitors are early B-type to late O-type stars that have exhausted their central hydrogen. Extending the Conti scenario \citep{Conti1975} to single massive stars in general, \citet{Chiosi1986} link the various spectral types into an evolutionary sequence.
\begin{center}
\scalebox{1.07}{%
\begin{tabular}{|llc|}
\hline
$M \gtrsim 60\, \msol$: & {\small O \donc Of/WNL \donc LBV \donc WNL \donc WC \donc WO} & \multirow{2}{*}{\large {\textbf{WR}}}\\
{$M \simeq 40-60\, \msol$}: & {\small O \donc BSG \donc LBV \donc WNL \donc (WNE) \donc WC \donc (WO)} & \\
\hline
\multicolumn{1}{l}{$M \simeq 30-40\, \msol$:} & {\small O \donc BSG \donc RSG \donc WNE\donc WCE} & \multicolumn{1}{c}{\null} \\
\hline
{$M \simeq 25-30\, \msol$}: & {\small O \donc (BSG) \donc RSG \donc (YSG? LBV?)} & \multirow{2}{*}{\large {\textbf{RSG}}}\\
{$M \simeq 10-25\, \msol$}: & {\small O/B \donc RSG \donc (Ceph. loop for $M \lesssim 15\, \msol$) \donc RSG} & \\
\hline
\end{tabular}
}
\end{center}
This historical scenario divides the O/B-type stars that become RSG and end their evolution as such, and the most massive O-types that become WR stars never expanding to the red side of the Hertzsprung-Russell diagram (HRD). In a small transitional mass range, O-type stars evolve to the red side of the HRD but lose enough mass there to evolve back to the blue and become WR (see Sect.~\ref{sec:blue_end}). Note that since then, our picture of massive star evolution has become more complex, and showed that the exact mass range for the different evolutionary paths depends on the physics considered. Moreover the Conti scenario does not take into account any binary interactions that would drastically modify these simple relations, for example allowing the appearance of low-luminosity WR stars, and changing the final type of supernova, favouring progenitors having been stripped of their envelope \citep[e.g.][]{Podsiadlowski1992,Vanbeveren2007,Eldridge2008,Eldridge2017,Li2024,Marchant2024}.

This scenario predicts that clusters should not host simultaneously RSG and WR populations coming from single stars, except in a very restricted age range (7-10 Myr). An overlap of both populations must come from binary interactions that created WR-like stripped stars through mass transfer.

\begin{figure}
\begin{center}
\includegraphics[width=0.9\textwidth]{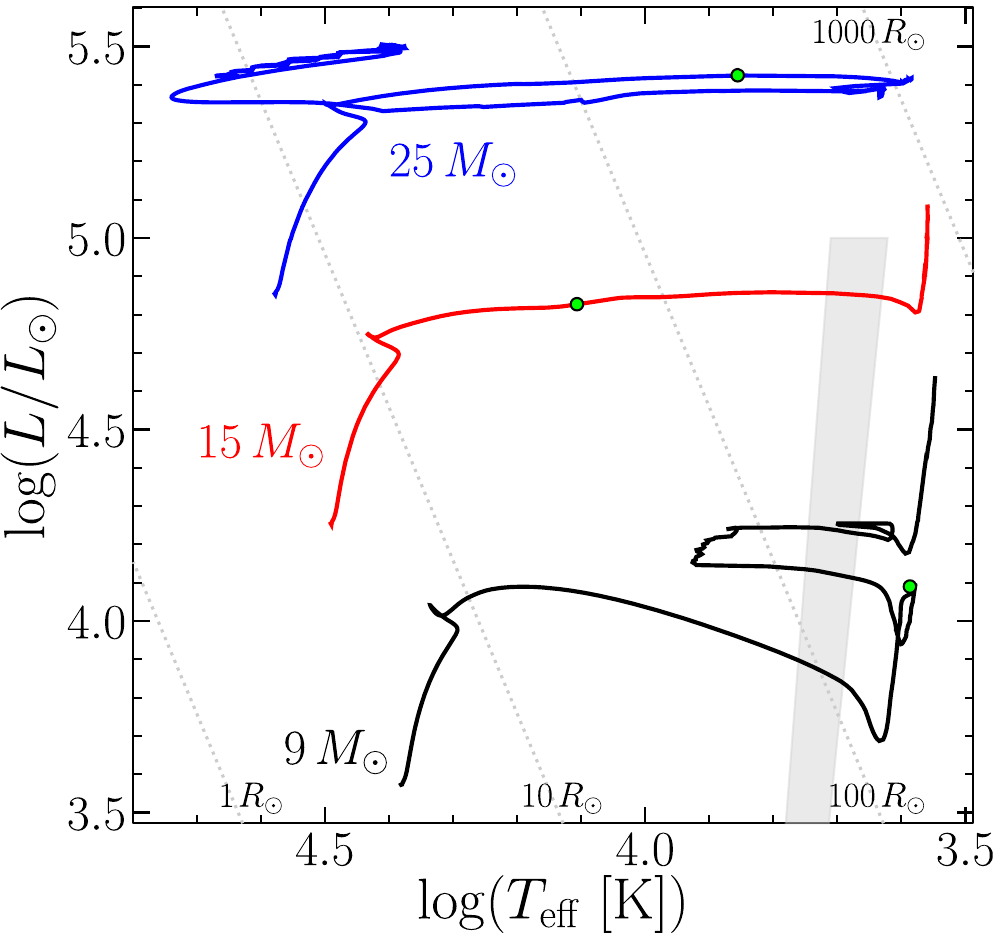}
\end{center}
\caption{HRD showing typical evolutionary pathway for three different initial masses at solar metallicity, and with an initial rotation $V_\text{ini} = 0.4\,V_\text{crit}$ (models from \cite{Ekstrom2012}). The green dots mark the location at the start of He burning. The 9\,{\msol} model (black) shows a quick crossing of the HRD at the end of the MS, and a blue loop during the RSG phase. The 15\,{\msol} model (red) shows a rather slow crossing to the RSG branch. The 25\,{\msol} model (blue) undergoes strong mass loss during the RSG phase, making the star to evolve bluewards at some point during the RSG phase, and end its life in the blue region of the HRD. The grey shaded area indicates the Cepheid instability strips, inside which a star is expected to have large radial oscillations (determination from \cite{Anderson2016}). Lines of iso-$R$ are displayed.}
\label{Fig:HRD}
\end{figure}

Assuming a Salpeter initial mass function (IMF,\cite{Salpeter1955}), then about 90\% of single massive stars will have an RSG phase at some point in their life\footnote{Here we assume it is the case for stars with an initial mass between 8\,{\msol} and 40\,{\msol}}, and 80\% of single massive stars would end their life while being a RSG.  These numbers highlight the importance of this phase for our understanding of massive star evolution, particularly for the advanced stages. According to the modified Conti scenario, we define three types of stellar pathways encompassing a RSG phase (see Fig.~\ref{Fig:HRD}):
\begin{enumerate}
    \item Stars that become RSG quickly after the end of the MS, and that undergo a blue loop before going back to the red and ending their life there;
    \item Stars that cross the {\hsg} and stay RSG until the end of their life;
    \item Stars that go to the RSG phase after the MS but evolve back to the blue and end their life there.
\end{enumerate}

It has been suggested that RSGs could be used as distance indicators for probing the Universe, either by looking at the brightest RSG of a galaxy \citep{Sandage1974,Glass1979} or by using a period-luminosity relation for these objects \citep{Jurcevic2000,Chatys2019}. However, the RSG luminosity seems to present a large intrinsic scatter, even when binarity effects are taken into account \cite{Wang2025}, so more work is needed to understand the origin of the scatter amplitude before the first method could be applied. For the second method, improvements of our understanding of the physics underlying the pulsations of RSGs are also needed in order to reach a reasonable efficiency in determining distances, particularly compared to other methods, such as the period-luminosity relation for Cepheid stars. 

With their high luminosity that gives the opportunity to see them in distant galaxies, RSGs have been proposed as reliable probes of metallicity even when only low-resolution IR spectroscopy is available \cite{Davies2010,Bergemann2012}.

RSGs are dust producers, the highest luminosity ones producing the largest grain dust \cite{Massey2005}. While in the present-day Universe, RSG are contributing for at most 10\% of the dust production \cite{Srinivasan2016}, massive Population III RSGs might have been important contributors in the early Universe \cite{Nozawa2014}, beside supernova explosions \citep{Slavin2020}, as requested by highly dusty galaxies at a redshift $>7$ \cite{Akins2023}. 

\section{Structure change after the Main sequence}
At the end of central hydrogen burning, the core is devoid of fuel and contracts. This contraction liberates gravitational energy that can be used to inflate the envelope. Often referred to as the {\it mirror effect}, the cause of this behaviour has been the subject of many studies \citep[see for instance][]{Yahil1985,Applegate1988,Bhaskar1991}. A simple explanation can be found in \citet{Padmanabhan2001}. The contraction of the core at the end of the MS occurs in a timescale that is of the order of the Kelvin-Helmholtz timescale $\tau_\text{KH}\simeq(GM^2/RL)$, but that is longer than the virialisation timescale. In this case, both the energy conservation ($U+\Omega=\text{const.}$) and the virial theorem ($2U+\Omega=0$) must hold true, with $U$ the thermal energy of the star and $\Omega$ its potential energy. The only way to achieve this is to have both $U$ and $\Omega$ conserved separately. Stars at the end of the MS have a larger part of their mass in the core than in the envelope ($M_\text{c}>>M_\text{env}$). The potential energy can be expressed as:
$$\left| \Omega \right| \approx \frac{GM_\text{c}^2}{R_\text{c}} + \frac{GM_\text{c}M_\text{env}}{R_\star}\approx \text{const.}$$
The location of the shell does not change significantly during the crossing of the {\hsg}, so we can consider that both $M_\text{c}$ and $M_\text{env}$ are constant:
$$-\frac{GM_\text{c}^2}{R_\text{c}^2}\frac{\mathrm{d}R_\text{c}}{\mathrm{d}t} - \frac{GM_\text{c}M_\text{env}}{R_\star^2}\frac{\mathrm{d}R_\star}{\mathrm{d}t}=0$$
We can then express the evolution of the star radius as a function of the core radius:
$$\frac{\mathrm{d}R_\star}{\mathrm{d}R_\text{c}}\approx -\left(\frac{M_\text{c}}{M_\text{env}}\right)\left(\frac{R_\star}{R_\text{c}}\right)^2$$
which shows how a contraction of the core triggers the expansion of the envelope.

The fact that stars cross the {\hsg} or not, and if so, the time needed to cross varies with stellar parameters. The strength of the intermediate convective zone (ICZ) building at the end of central H burning, on top of the H-burning shell, and the chemical gradient above it plays a crucial role \cite{Sugimoto2000}. When the shell is strong, with a high energy generation, it supports the star and slows the crossing \cite{Sibony2023}. Any condition leading to a deep and/or active shell favours a large ICZ. This is the case for larger initial masses, where the shell builds in hotter conditions: for stars more massive than 12\,\msol, He starts burning while the star is still crossing the {\hsg} because the shell's sustain is efficient enough to prevent the star to quickly cross the {\hsg}. Similarly, a low metallicity slows the crossing down, because the compactness of the star \cite{Maeder2001} locates the shell in a hotter region than in a star at solar metallicity.

For the same reasons, there is a dependency on the criterion used for the definition of convective zones. When the Ledoux criterion ($\nabla_\text{rad}>\nabla_\text{ad}+\frac{\varphi}{\delta}\nabla_\mu$, \cite{Ledoux1947})\footnote{with $\nabla_\text{ad}=\left(\frac{\partial\ln T}{\partial\ln P}\right)_\text{ad}=\frac{P\delta}{C_P\rho T}$ (where $\delta$ comes from a general equation of state of the form $\Delta\ln\rho = \alpha\,\Delta\ln P - \delta\,\Delta\ln T + \varphi\,\Delta\ln\mu$), the radiative gradient $\nabla_\text{rad}=\frac{3}{16\pi a c G}\frac{\kappa LP}{MT^4}$ (where $\kappa$ the opacity and obvious meanings for $L$, $P$, $M$, and $T$), and $\nabla_\mu$ the gradient of the average mean molecular weight.} is taken instead of the Schwartzschild one ($\nabla_\text{rad}>\nabla_\text{ad}$, \cite{Schwarzschild1958}), the chemical gradient just above the initial location of the H-burning core prevents convection, so the ICZ is located higher in the star and is weaker. The crossing is quick and accompanied by a drop in luminosity, with part of the radiation being used to inflate the envelope.

\begin{figure}
    \centering
    \includegraphics[width=\linewidth]{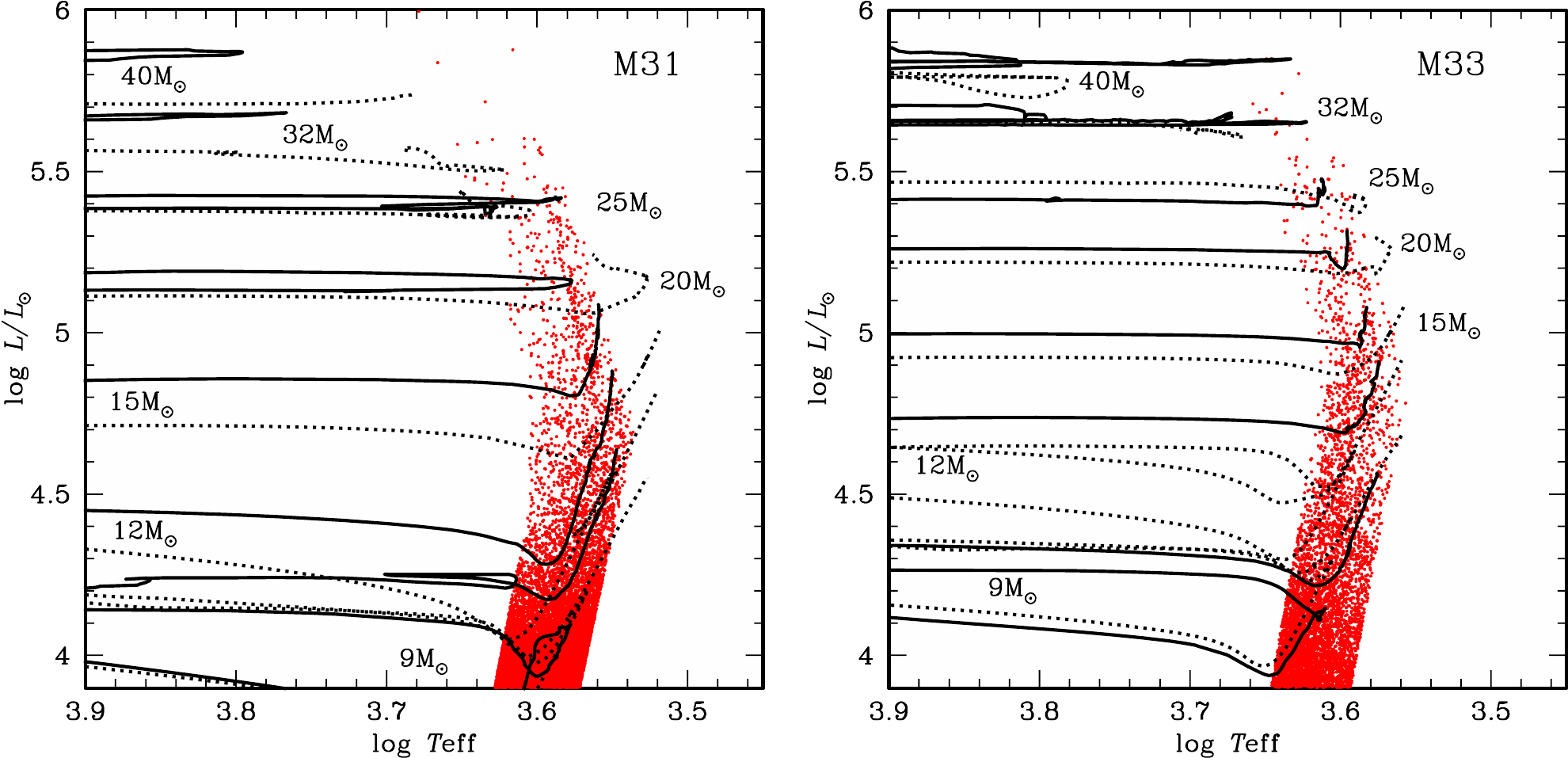}
    \caption{HRD of observed RSG (\cite{Massey2021}, red dots) in M31 (\textit{left}), and M33 (\textit{right}). Stellar evolution tracks are overplotted: GENEC models (solid lines) from \cite{Ekstrom2012} for M31 ($Z=0.014$), and from \cite{Eggenberger2021} for M33 ($Z=0.006$); MIST models (dotted lines) from \cite{Choi2016} with [Fe/H]=0 for M31, and [Fe/H]=-0.5 for M33. Adapted from Fig.~8 of \cite{Massey2021}.}
    \label{fig:Massey_RSG}
\end{figure}

Another reason to cross the {\hsg} quickly is linked to the overall opacity of the envelope, and hence to the mass of the H-rich envelope on top of the shell. The larger the mass of the envelope, the faster the crossing takes place \cite{Farrell2020a}. This imposes a natural limit to the maximal mass of RSG: above $\sim 30\,\msol$, the MS mass loss removes enough of the H-rich envelope for the star to remain in the blue side of the Hertzsprung-Russel diagram (HRD). This mass limit translates into a luminosity limit when we compare the models to observations (see Fig.~\ref{fig:Massey_RSG}) At low metallicity, the MS mass loss is weaker, but the stars are maintained in the blue by the Humphreys-Davidson limit \cite{Humphreys1979} and the inherent mass loss triggered by instabilities. Thus, the highest luminosity that RSGs can reach does not seem to depend on metallicity \cite{Davies2018}.

\section{Structure and evolution in the RSG phase}

The RSG phase is marked by a strong contrast between the contracted central parts reaching the high temperature of He burning, and the very extended envelope cooling down as it expands. The opacity increases strongly because of the contribution of metal lines and, for the coolest RSGs, of molecules. The energy produced in the core cannot be transported by radiation only through the envelope, so convection is triggered, and a deep outer convective zone is built.

\subsection{A structure dominated by convection}
\begin{figure}
    \centering
    \includegraphics[width=\linewidth]{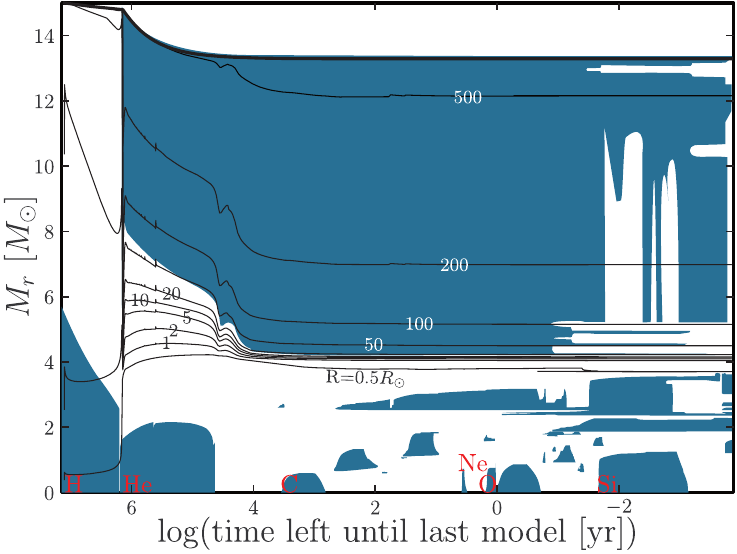}
    \caption{Kippenhahn diagram of a 15\,{\msol} model at solar metallicity (model from \cite{Griffiths2025}). The convective zones are the blue shaded regions. Lines of iso-$R$ are displayed. The central burning phases are indicated in red. The thick black line shows the total mass evolution of the star.}
    \label{fig:kipp15}
\end{figure}

The outer convective zone dives deeply inside the stellar interior, engulfing the outer 60\%-70\% of the total mass\footnote{\footnotesize{It is even more impressive when expressed in terms of the radius of the star. The outer convective zone of a RSG covers more than $99\%$ of the total radius of the star.}} (see Fig.~\ref{fig:kipp15}). This means that it reaches a zone where H burning took place previously, where the chemical composition has been modified by nuclear reactions (in particular, C and O have been transformed into N). The convective movements transport this material up to the surface \cite{Iben1964}, providing a chemical enrichment even to non-rotating stars that are not supposed to mix otherwise. This is known as the First Dredge Up. The effect of this dredge up on the surface abundances depends on the exact modelling of the star. However, the general trend is that CNO-burning products should appear at the surface, i.e. an increase in the nitrogen abundance, and a depletion of carbon and oxygen. This seems to be confirmed by observations of RSG in the Galaxy \citep{Davies2009a,Davies2009b}.

The surface of RSGs is sculpted by convection. Direct imaging \cite{Gilliland1996} or interferometric observations of Betelgeuse and other RSGs \cite{Tatebe2007,Hautbois2009} show inhomogeneities and aspheric structures that are attributed to convective granulation. Radiative hydrodynamic simulations confirm that convective cells are large and strongly asymmetric \cite{Chiavassa2009,Chiavassa2010}. The characteristic size of convective granulation ($x_g$) is expected to scale with the atmospheric pressure scale height $H_{P0}$ like $x_g=\alpha\,H_{P0}$, with $\alpha$ around 10 for AGB stars \cite{Freytag1997}. As found by \cite{Chiavassa2009}, in the case of RSGs, the effect of the turbulent pressure has to be taken into account since in their simulations, the turbulent pressure is larger than the gas pressure by a factor of 2. Using the definition of the turbulent pressure from \cite{Gustafsson2008}, they obtain the pressure scale height $H_{P0,\text{turb}}=\frac{k\teff}{g\mu m_\text{H}}\,\left(1+\beta\gamma\left(\frac{V_\text{turb}}{c_\text{s}}\right)^2\right)$, with $\beta$ a factor close to one, $\gamma$ the adiabatic exponent, and $c_\text{s}$ the sound speed. This results in a granulation size approximately five times larger than when turbulent pressure is neglected. The largest convective cells evolve on a time scale of years, whereas smaller features vary on a time scale of months \cite{Norris2021}. At the surface of RSG stars, radiation-hydrodynamics simulations show that the contrast between the brightest and the darkest parts can reach a factor of about $50$ \citep{Chiavassa2010}.

The variability induced by the convective movement on the surface of RSG contributes to a noise that blurs the determination of the position by astrometric measurements \cite{Chiavassa2011}. For a Betelgeuse-like RSG, typically, the photocentre moves randomly with an amplitude of the order of 0.1\,AU, significantly impacting the determination of the parallax, and hence the distance. Conversely, by measuring the dispersion in parallax measurements of RSGs that are members of a cluster (for which the distance can be known thanks to the other members), constraints on the surface convective dynamics could be obtained \cite{Chiavassa2022}. Unfortunately it seems that the sensitivity of current astrometric instruments like \textit{Gaia} is still about an order of magnitude too low to be able to give meaningful results \cite{Kochanek2023}. Convection also contributes in producing photometric variability, as shown by multi-D radiation-hydrodynamics simulations \citep[e.g.][]{Chiavassa2011}, reaching order of a few tenths of dex in magnitude. Luminosity variations attributed to the interplay between convection and pulsations have also been observed in long-term monitoring of RSG sample \citep{Kiss2006}.

\subsection{Radius increase and binary interactions}
On the main sequence, the precursors of RSGs are O- and B-type stars, which are notably known to have a binary occurrence fraction larger than 60\% \cite{Sana2013}. However, the binary occurrence fraction of RSGs is only around 30\% \cite{Patrick2020,Neugent2021,Dai2025}, dropping below 20\% at low metallicity \cite{Patrick2020,Neugent2021,Patrick2022}. The "missing binaries" have several explanations. Typically, O- and B-type stars have radii of about 10\,{\rsol}, but during the crossing of the Hertzsprung gap, the radius increases enormously, up to about 1500\,{\rsol} for the largest known RSGs (see e.g. \cite{Wittkowski2012}). With such an increase in radius, all systems with an orbital period lower than $\sim$1500\,days are expected to undergo binary interactions, such as mass transfer, common envelope evolution, or merger, either already on the MS for the closest ones, or when the primary leaves the MS \cite{Patrick2020}.

In the case of the closest systems ($P\lesssim 10$\,days), it is considered that there is a high probability that the two components merge already during the MS, creating a blue straggler. Some of them can later evolve into a red straggler, a RSG too luminous compared to its native cluster (so
inferred/misinterpreted as too young, \cite{Britavskiy2019}) and that shows no sign of binarity any more. Generally, mergers of close systems occurring during the MS can lead to isolated supergiants that will increase the RSG luminosity dispersion of clusters \cite{Wang2025}. The merger can also occur later, after a common-envelope phase triggered by the post-MS inflation of the primary, like the probable scenario for the progenitor of SN1987A \cite{Podsiadlowski1990} and of type II SN in general \cite{Podsiadlowski1992,Zapartas2019,Schneider2024}. 

For periods between 10 and 1500\,days\footnote{The exact values depend on physical assumptions made in the models.}, the expected binary interactions would lead the primary to become a stripped star, and if the mass transfer is stable, only the secondary would evolve to the RSG. In this case, it would either show signs of a compact companion, or the explosion of the primary would have disrupted the binary, the RSG appearing as a single, run-away star (or rather walk-away \cite{Renzo2019}), like $\alpha$\,Ori \cite{Harper2008}.

For longer periods ($P>1500$\,days), both stars essentially evolve as in isolation, preserving their binary nature in the RSG phase. If the RSG is the primary component of the binary system, and if the companion is a B-type star (which seems to be a very common situation \cite{PantaleoniGonzalez2020}), it can be detected by a photometric blue excess \cite{Neugent2021,Patrick2022}. Radial velocities for such long periods typically range between 1-5 km/s stretched on timescales of years or decades \cite{Patrick2020}. Since these amplitude and period are close to those coming from the convective movements at the surface of the RSG, the binary diagnostic through radial velocities is a difficult one.

\subsection{Mass-loss regime}
In general, the mass-loss regime shifts as a star enters the RSG phase  (see the review by van
Loon in this Special Issue). Despite the high opacity due to the low temperature, a radiatively-driven wind regime does not appear to be applicable to RSGs. Convective plumes can rise high above the "surface" \cite{LopezAriste2023} and could be at the origin of strong mass-loss episodes \cite{Smith2001, Montarges2021, Humphreys2022}. Beside the uncertainties in the mechanism(s) driving the mass loss, binarity adds a potential source of mass loss through transfer on the companion. The mass lost during the RSG phase can be probed either directly by measuring current mass-loss rates of stars, or inferred indirectly by the study of their effects on stellar populations. Direct estimates of RSG mass-loss rates are abundant in the literature, though often contradictory, varying by orders of magnitude for a given RSG luminosity \cite{Mauron2011}. When a restricted mass domain and parameter space is targetted, as in the well-selected clusters determination performed by \cite{Beasor2018}, a tight correlation is found between the luminosity of the RSG and the mass-loss rate. In this work, the authors find lower mass-loss rates than the ones commonly used in stellar evolution codes.  Recent works \cite{Humphreys2020,Yang2023, Antoniadis2024} find a kink in the $\dot{M}-L$ relation, with the slope steepening above a given luminosity ($L/{\lsol} \gtrsim 4.3$). Indirect methods are, for example, the radio signature of SNe \cite{Moriya2021} or the luminosity function of RSG \cite{Massey2023}. Their results seem to not favour low rates nor very high rates, although it is difficult to draw a firm answer and disentangle what comes from stellar mechanisms and what comes from binary interactions. An still open question is the metallicity dependence of the RSG mass-loss rates. While measuring the $\dot{M}-L$ relation in the Magellanic Clouds is doable, that in the Milky Way is more complicated due to uncertainties in distances and reddening \cite{Antoniadis2025}.

\begin{figure}
    \centering
    \includegraphics[width=.9\linewidth]{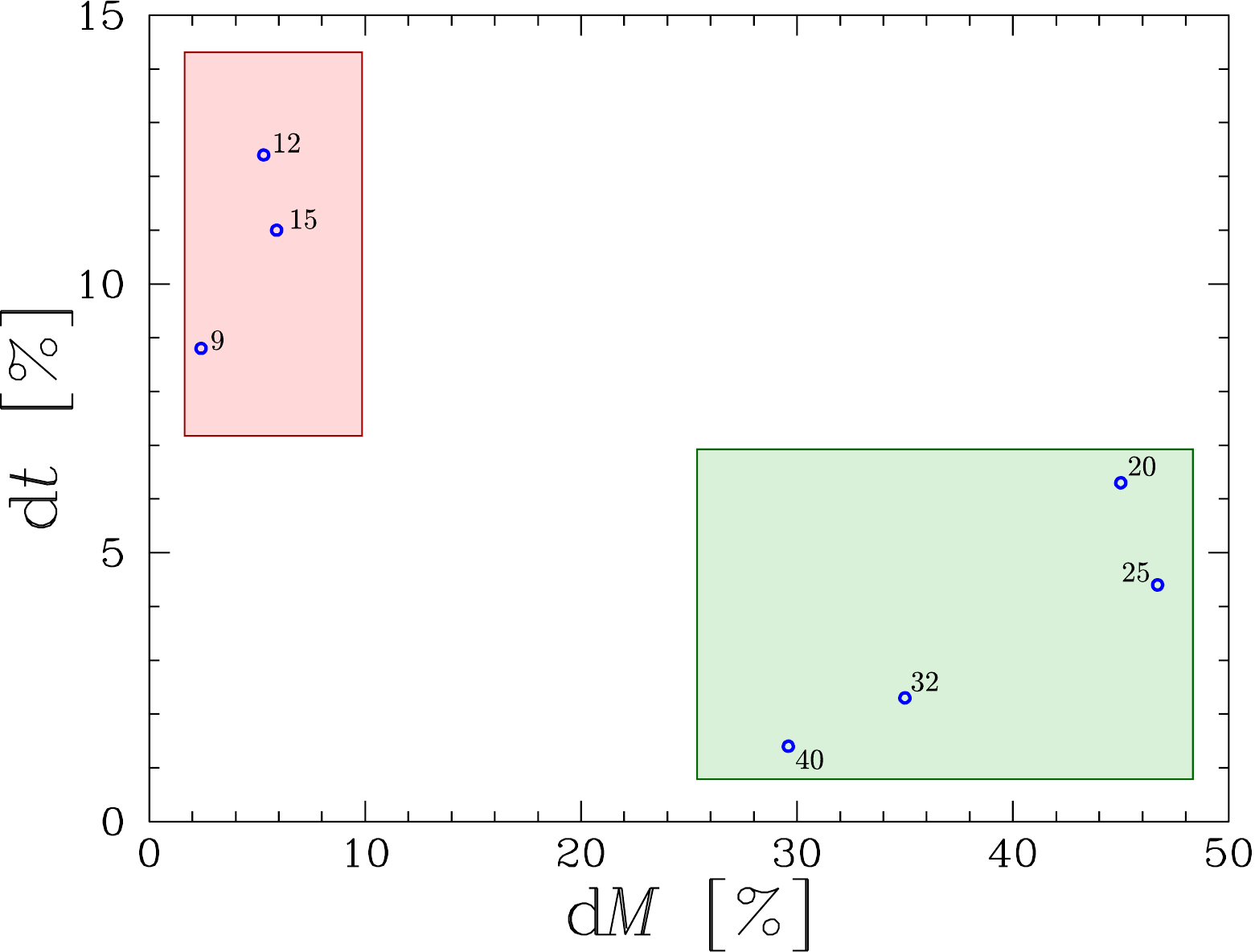}
    \caption{Duration of the RSG phase (in fraction of the total lifetime) as a function of the mass lost in this phase (in fraction of the total initial mass). Data from the non-rotating models of \cite{Ekstrom2012}. The region of low d$M$ - large d$t$ (occupied by the lowest-mass models) is shaded in red, while the region of large d$M$ - low d$t$ (occupied by the highest-mass models) is shaded in green.}
    \label{fig:dMdt}
\end{figure}

When observed mass-loss rates are translated into a recipe for stellar modelling, the main difficulty resides in the timesteps that are used by the models. Typical timesteps are of the order of decades or centuries, and they need to capture and average the bursty nature of RSG winds. The magnitude of the mass-loss rates has strong consequences in the predictions of stellar evolution in the RSG phase as well as endpoints or supernovae types \cite{Zapartas2025}. A strong mass loss during the RSG (be it from high mass-loss rates, from outbursts events \cite{Cheng2024,MunozSanchez2024a}, or from a mass transfer in binaries), is supposed to lead the star to a blueward evolution (\cite{MunozSanchez2024b}, also see Sec.~\ref{sec:blue}). Models show a clear division between the ones that lose a lot of mass and stay a short time in the RSG region, and those losing less mass and spending a long time in the RSG region (see Fig.~\ref{fig:dMdt}).

\subsection{Late stages}
After central helium burning exhaustion, the core contracts and heats, reaching temperatures where carbon burning can start. From this stage on, the energy produced inside the star is mostly evacuated by the mean of neutrinos rather than photons, and escapes the star without interacting with the gas. As a simple, back-of-the-envelope estimation, the nuclear timescale of the C-burning phase, which is the longest of the advanced phases, can be expressed as (contraction + C burning): $$\tau_\text{contr} + \tau_\text{C-b} = \frac{\frac{1}{2}\Delta\Omega + E_\text{C-b}M_\text{c}}{L_\nu} \sim 1000\,\text{yr},$$ where $\Delta\Omega$ is the gravitational energy liberated by the contraction, $E_\text{C-b}$ the specific energy produced by C burning, $M_\text{c}$ the mass fraction of the core where the burning occurs, and $L_\gamma$ the luminosity emitted in the form of neutrinos. It becomes shorter than the Kelvin-Helmholtz time scale on which the envelope evolves: $$\tau_\text{KH} = \frac{GM^2}{RL_\gamma} \sim 10^5-10^6\,\text{yr}.$$ The envelope therefore has no time to react to whatever happens in the centre of the star: the external appearance of the star is no longer changing, even if major changes occur in the central region, while multiple burning and contraction episodes succeed each other.

However, very early SNe observations reveal the latest events of the stars life, and offer a glimpse into the closest region of the circumstellar medium (CSM, \cite{Morozova2018}). In some cases of RSG progenitors, they point to a strongly increased mass loss shortly prior to the explosion. A steady wind lasting for more than a few decades is ruled out by pre-explosion images, because such a wind would veil the progenitor in dust and alter its appearance. Only an outburst or a strong wind lasting at most a few years is supported by early SN analyses \cite{Davies2022, Hiramatsu2023}. Curiously, this CSM layer of thick and slow wind around ready-to-collapse RSG can mimic a WR wind and its typical HeII $\lambda$4686 and CIII/NIII features \cite{Smith2015}.In some other cases, such as SN  2023ixf \citep{Kilpatrick2023} or SN 2024ggi \citep{Xiang2024}, the progenitor was identified as a classical RSG without indication of strong outburst prior to the explosion.

\section{Blueward evolution and loops \label{sec:blue}}
While some stars remain RSGs once they enter this phase, others may either temporarily exit the RSG regime or permanently evolve back towards the blue. The former scenario is associated with Cepheid blue loops, whereas the latter pertains to the most massive RSGs.

\subsection{Cepheids and blue loops}
As they evolve along the red (super-)giant branch, stars with masses below approximately 12\,{\msol} undergo a temporary blueward excursion before returning to the red and ultimately ending their lives as RSGs (see the 9\,{\msol} track in Fig.~\ref{Fig:HRD}). During this blue excursion and back, the stars cross a region that is known as the {\it Cepheid instability strip} where they undergo pulsations \cite{Sandage1971,Gautschy1995,Bono1999}. These blue loops develop progressively at the lower end of the mass range, with their extent reaching bluer regions of the HRD for more massive stars. At the upper end of this range, a sharp transition occurs from a maximally extended loop to the complete absence of a loop \cite{Anderson2016}.

The mechanism responsible for the launch of a blue loop is extremely complex. As expressed in Kippenhahn \& Weigert: "\textit{We see that details, which have originated from different regions and from earlier phases when the effects were scarcely recognizable, can now pop up and modify the evolution appreciably. The present phase is a sort of magnifying glass, also revealing relentlessly the faults of calculations of earlier phases.}" \cite{Kippenhahn1990}. Blue loops have been the subject of many studies. Parametric, static models with 2 or 3 zones have been used in the 1970s to explore under which conditions the solution of the stellar equations results in a blue or red location \cite{Lauterborn1971, Fricke1972, Schlesinger1977}, but causes and effects are difficult to disentangle. The key factor influencing the appearance or not of a loop seems to be the helium excess above the H-burning shell, a large composition discontinuity having a suppressing effect on the looping behaviour: while early studies evoked the potential of the core $\sim \frac{M_\text{c}}{R_\text{c}}$ as an influencing factor for the loops, it seems that it is not so much its value that plays a role, but rather its influence on the position of the H-burning shell and hence the H-He discontinuity \cite{Walmswell2015}. Any process that would modify the steepness of the discontinuity or the depth at which the outer convective zone dives will have an influence on the loops, either suppressing them or changing their extent \cite{Tang2014,ZhaoL2023}.

Since the launching of a loop is triggered by the H-burning shell reaching the composition discontinuity left by the first dredge-up, it usually takes half the He-burning phase to reach that point, and the blue loop occurs in the second half of the He-burning phase. The end of the loop takes place very shortly before He exhaustion in the core. After the loop, the star moves back on the RSG branch and ends as a type II SN.

Given the sensitivity of the blue loops on the depth of the convective dredge-up and on the distance between the H-burning shell and the composition discontinuity, the mass range at which a blue loop is expected changes with metallicity. At solar $Z$, blue loops are expected from $2.5-12\,\msol$ (but for the loop to reach the Cepheid instability strip, only $M>4\,\msol$ do so). At low $Z$, the mass range is slightly shifted down. The loops are wider, and the most massive stars climb on the loop before having fully reached the RSG region \cite{ZhaoL2023}.

\subsection{Blueward excursion as the final evolution\label{sec:blue_end}}

Mass loss becomes an important process during the RSG phase (see the review by van Loon in this Special Issue). Since it is usually believed that the mass-loss rates scale with luminosity \citep{deJager1988,Beasor2020}, they can reach high values for more luminous RSGs near the end of their life. The mass-loss rates during the RSG phase are relatively poorly known and vary a lot depending on the authors, the sample used, the type of stars, among other factors \citep[e.g.][see also \citealt{Meynet2015}]{vanLoon2005,Mauron2011,Beasor2020}. In case the total mass lost during this phase is large enough to decrease significantly the mass of the H-rich envelope, it will make the star leave the RSG branch and evolve bluewards in the HRD, becoming a yellow hypergiant (see the review by Jones \& Humphreys in this Special Issue), or possibly becoming a WR star if it were to lose its envelope almost completely. The blueward evolution of the RSG usually starts when the mass of the core exceeds about 60\% of the total stellar mass \citep{Giannone1967}. Depending on the mass-loss prescription used, this can occur for stars with an initial mass around $15\,\msol$ \citep{Georgy2012, Meynet2015}, or not at all below about $30\,\msol$ \citep{Beasor2021}. Knowing precisely the mass-loss rates of RSG is thus of prime importance for stellar evolution, not only during the quiescent phases but also in the case of less common events of large outbursts, that can produce quick evolution of the effective temperature (and therefore colour) of the star on very short timescales \citep{MunozSanchez2024b}.  If the total mass lost during this phase remains relatively modest, the star will remain a RSG until the end of its life, leading to a type IIP supernova \citep{Beasor2021} or directly collapse into a BH \cite{Fryer1999}. On the other hand, if more copious amounts of mass can be lost during the RSG phase, the star will evolve further as a yellow supergiant, and then possibly as a blue supergiant or even a WR star \citep{Georgy2012b}. This would allow stars to end their life at different locations in the HRD, producing different types of SN events: intermediate type IIL or IIb for stars with an end-point in the yellow part of the HRD \citep{Georgy2012}, or type Ib or even type Ic for more massive stars able to completely remove their H-rich envelope \citep{Georgy2012b, Meynet2015} (see also the review by Van Dyk in this Special Issue). In case the RSG star has a close enough companion, strong mass loss from the RSG is also possible through stripping by binary interactions, leading to similar scenarios, possibly at lower mass than in a single star. Knowing precisely the mass-loss history during the RSG phase is thus mandatory for a correct modelling of the late stage of massive stars evolution, including the occurrence of strong outbursts \citep{Cheng2024}. Our lack of understanding of the mass-loss mechanism in such stars leads to major uncertainties in their further evolution \citep{Zapartas2025}.

Observationally, the idea that mass-loss rates of RSGs could have been underestimated has emerged from the so-called ``red supergiant problem''. In fact, it appears from the study of the progenitors of a sample of type IIP supernova that no one of them had an initial mass greater than about $16.5\,\msol$ \citep{Smartt2009,Smartt2015}. This implies that either more massive RSGs directly collapse into a BH \cite{Fryer1999}, or they evolve towards other regions of the HRD \cite{MunozSanchez2024b} prior to explode into another supernova type. In case of a direct collapse into a BH, a massive star would simply disappear from the sky, with or without a faint outburst \cite{Kochanek2008,Lovegrove2013}. Tentative searches for such a disappearance have been conducted in archival images (\cite{Reynolds2015}) or dedicated surveys (\cite{Gerke2015}), yielding a few robust candidates, and compatible with the fraction of failed supernova of 0.1--0.4 expected from the observed mass function of BH \cite{Kochanek2015}. On the other hand, losing high-$L$ RSG through a late excursion to the blue is being possible by increasing the mass-loss prescription in stellar evolution codes or by considering late mass-transfer episodes. However, note that this finding has been debated recently \citep{Davies2020, Farrell2020b}, since it would require mass-loss rates during the RSG phase considerably higher than those recently determined \citep[e.g.][]{Beasor2018,Beasor2020}.

Another way of verifying whether a blue supergiant star is pre- or post-RSG phase is to compare some of its surface properties with observations. It has been shown that the surface abundances and pulsational properties of variable blue supergiant are best in line with stellar evolution prediction of post-RSG evolution for such stars, indicating that they are possibly on a bluewards evolution due to mass loss \citep{Saio2013,Georgy2014}. Indeed, a strong mass loss (by winds or through binary interaction) during the RSG phase considerably increases the $L/M$ ratio of these stars, greatly favouring the triggering of pulsations, making their luminosity vary over time.

\section{Conclusions and perspectives}
As discussed above, the majority of massive stars will undergo an RSG phase. It is therefore of prime importance to have a good understanding of this phase, in order to have a better comprehension of massive star evolution in general, of the type II P supernovae progenitors\footnote{\footnotesize{We recall here that type II P SNe is by far the most numerous type of core-collapse supernovae \citep{Smith2011}.}}, and of the precursors of various types of massive stars in their late stages. Seen from the point of view of stellar modelling, the most critical process to be constrained is the mass loss rates of RSGs. Not only the regular mass loss, but also possible mass-loss events related to eruptions, where a consequent quantity of mass can be lost in a short timescale. How frequent are these events? How much mass do they remove? How does metallicity affect the mass-loss budget of RSGs? Having strong constraints on this would definitely help in better modelling the RSG stage and the possible subsequent evolution of massive stars. Given their importance, all these questions are currently the subject of active cutting-edge research.

\vspace{6pt} 

\acknowledgments{SE acknowledges support from the Swiss National Science Foundation (SNSF), grant number 212143. The authors are grateful to the anonymous referees who helped them improve this review.}

\abbreviations{Abbreviations}{
The following abbreviations are used in this manuscript:\\

\noindent 
\begin{tabular}{@{}ll}
BH(s) & Black hole(s) \\
CSM & Circumstellar medium \\
HRD & Hertzsprung-Russell diagram \\
ICZ & Intermediate convective zone \\
IMF & Initial mass function \\
MS & Main sequence\\
RSG(s) & Red Supergiant(s)\\
SN(e) & Supernova(e) \\
WR & Wolf-Rayet
\end{tabular}
}

\begin{adjustwidth}{-\extralength}{0cm}

\reftitle{References}


\bibliography{RSG_evol_Ekstrom_Georgy}

\PublishersNote{}
\end{adjustwidth}
\end{document}